\documentclass[superscriptaddress,showkeys,showpacs,preprintnumbers]{revtex4}
\usepackage{feynmf}
\usepackage{graphicx}
\usepackage{color}

\usepackage[tbtags]{amsmath}
\newcommand{\mjet}{\mbox{\tiny jet}}
\newcommand{\mojet}{\mbox{\tiny 1 jet}}
\allowdisplaybreaks[4]
\begin{document}
\begin{fmffile}{fmfp}
\title{ ${\cal{O}}(\alpha_s^2)$ QCD corrections to inclusive jet production in 
deep inelastic scattering.}
\thanks{Partially supported by CONICET, Fundaci\'on Antorchas, UBACYT and 
ANPCyT, Argentina.}
\author{A. Daleo}
\email{daleo@physik.unizh.ch}
\affiliation{Institut f\"ur Theoretische Physik,  Universit\"at Z\"urich, \\
Winterthurerstrasse 190, CH-8057 Z\"urich, Switzerland}.
\author{R. Sassot}
\email{sassot@df.uba.ar}
\affiliation{Departamento de F\'{\i}sica,
Universidad de Buenos Aires\\ Ciudad Universitaria, Pab.1 (1428)
Buenos Aires, Argentina}
\preprint{ZU-TH 21/05}

\date{{\bf \today}}

\begin{abstract}
We analyze the order $\alpha_s^2$ corrections to the single inclusive
jet cross section in lepton-nucleon deep inelastic scattering. 
The full calculation is done analytically, in the small cone approximation, 
obtaining finite NLO partonic level cross sections for these processes.
A detailed study of the different underlying partonic reactions is presented
focusing in the size of the corrections they get at NLO accuracy, 
their relative weight, and the residual scale uncertainty they leave in 
the full cross section depending on the kinematical region explored. 
The dominant partonic process in very forward jet 
production is found to start at order $\alpha_s^2$, being effectively a 
lowest order estimate, with the consequent large factorization scale 
uncertainty, and the likelihood of non-negligible corrections at the 
subsequent order in perturbation theory.  
\end{abstract}


\maketitle

\section*{Introduction}
Over the last thirty years, the DGLAP \cite{DGLAP} approach to parton dynamics 
has demonstrated itself as the most adequate tool for the description of the 
energy scale dependence of a variety of lepton-nucleon, and nucleon-nucleon 
processes over a wide kinematical range.
Surprisingly, not just this approximation, but the lowest order in 
perturbation theory within this approach (LO) gives fairly accurate estimates 
for 
paradigmatic processes such as inclusive deep inelastic scattering (DIS), 
provided an energy or momentum scale of a few GeV characterizes them. 
The following order (NLO) often represents small corrections, required for 
precise comparisons, but not for the broad picture.

In the few last years high precision DIS experiments, with a wide kinematical 
coverage, and the ability to measure less inclusive processes, such as those 
performed by the ZEUS and H1 collaborations at HERA, have extended the tests 
on the dynamics of partons to the limits of their kinematical reach, looking 
for signatures of 
dynamics complementary to that described by the DGLAP approach. Illustrative 
examples of these tests are the measurements of final state hadrons 
\cite{Aktas:2004rb} and jets \cite{jetsZ,jetsH1} produced in DIS processes 
in the forward region, for which the LO DGLAP description 
fail to reproduce the data by an order of magnitude, and even NLO estimates 
fall short.

In a recent analysis \cite{Daleo:2004pn}, it has been shown that the striking 
failure of the LO description in the case of forward hadrons by no means 
implies the breakdown of the DGLAP dynamics, but just the inadequacy of the LO 
picture, which simply does not include the dominant contribution to the 
measured cross section: the process in which an initial state gluon is knocked 
out from the nucleon, and also a gluon fragments into the detected final state 
hadron. Indeed, the NLO approximation, which takes into account these 
contributions, reproduce nicely the data 
\cite{Daleo:2004pn,Kniehl:2004hf,aure}. 
 
In the case of forward jets, the situation seems to be more 
compromised, because not only the LO estimates fail, but also NLO estimates
fall short by a factor of two of the data \cite{jetsH1}. In reference 
\cite{jetsZ} this feature together with the large scale dependence of NLO 
calculations, has been taken as indicative of the importance of higher order 
corrections. In order to improve the understanding of 
this situation, in the present paper we compute the order $\alpha_s^2$ 
corrections to the single inclusive jet cross section in lepton-nucleon deep 
inelastic scattering. 

We perform the full calculation analytically, factoring out explicitly the 
remnant initial state singularities into the definition of parton densities,
and dealing with those of final states in the small cone approximation, which 
assumes that the cone of the jet is narrow \cite{SCA}.  This approximation 
allows 
us to translate straightforwardly previous results on hadroproduction 
\cite{Daleo:2004pn} to the 
case of jets, avoiding a rather cumbersome calculation and delicate numerical 
treatments for dealing with the collinear singularities. In Section I we 
outline the technical framework required for the small cone approximation 
(SCA) in the case of DIS. 

The small cone technique approximates the full result obtained 
by NLO Monte Carlo generators, providing fairly good estimates for the cross 
sections with differences typically smaller than the theoretical 
uncertainties of the NLO estimate in the relevant kinematical range. We 
investigate in section II the range of validity of the small cone approximation 
against the full Monte Carlo NLO parton generators.

The finite cross sections obtained within the small cone approximation can 
be implemented in fast and stable codes, and simple to use. Exploiting the 
flexibility of them, in Section III we perform a detailed study of the 
different underlying partonic reactions and their main features: the size of 
the corrections they get at NLO accuracy, their relative weight, and the 
residual scale uncertainty they leave in the full cross section depending on 
the kinematical region explored.              

Our main conclusion is that, as in the case of hadroproduction, the dominant 
partonic process in the most forward jet region accessed yet is the one with 
a gluon in the initial state and also a gluon as the main seed of the jet. 
This process starts at order $\alpha_s^2$, and makes the NLO effectively a 
lowest order estimate, with the consequent large factorization scale 
uncertainty, and the likelihood of non-negligible corrections at the 
subsequent order in perturbation.

\section{Single-jet-inclusive DIS cross section in the SCA.}

In this section we outline the calculation of the single-jet-inclusive DIS 
cross section within the small cone approximation. This technique has 
been proposed \cite{SCA} and used in computations of unpolarized single 
inclusive cross sections \cite{SCA2}, and more recently has been extended 
and validated for polarized proton-proton collisions \cite{Jager:2004jh}. 

The SCA can be thought as an expansion of the partonic cross section in 
terms of the half aperture $\delta$ of the cone over which the radiation 
around a given final state parton is integrated. This cone defines the jet 
at partonic level and the integration over it regularizes all final state 
collinear singularities. Keeping the most significant terms in this 
expansion, which is given by $a\,log(\delta)+b+{\cal{O}}(\delta^2)$ 
and neglecting ${\cal{O}}(\delta^2)$ and higher contributions, the expansion
approximates surprisingly well the full results of the cone algorithm 
for cone radius up to $R\simeq 0.7$ in applications for proton-proton 
collisions \cite{Jager:2004jh}.

The main advantage of the approach in the present case is that it is possible 
to relate analytical results on one-particle-inclusive cross sections 
calculated previously \cite{Daleo:2004pn}, where all collinear divergences 
have already been canceled or factorized into parton densities and 
fragmentation functions, to the single-jet-inclusive cross section, for which 
the SCA gives an analytical expression as a function of the cone aperture. 
As in reference \cite{Daleo:2004pn}, we restrict the discussion to the case of
non-vanishing transverse momentum. The limit of vanishing transverse momentum
requires a more involved treatment of collinear singularities 
\cite{Daleo:2003jf,Daleo:2003xg}.
Schematically, we start with the finite one-particle-inclusive cross section, 
we undo the factorization of final state collinear singularities into 
fragmentation functions, which of course is not pertinent in the case of jet 
cross sections. Then, we add contributions that are not accounted for in the 
one-particle-inclusive fragmentation scheme, i.e. jets formed by two partons. 
The result is completely finite and  
for phenomenological applications can be convoluted 
with parton densities just as in the case of the one particle inclusive cross 
section. In practice, the approach is 
equivalent to having defined effective jet fragmentation functions that 
factorize the final state singularities.

\subsection{Kinematics}

In order to implement the approach in the case of DIS, we start defining the 
total cross section for the partonic process 
$a+l\rightarrow \mbox{1 Jet}+l^\prime+X$ with $n$ partons in the final state: 
\begin{equation}\label{eq:totalcs}
\frac{d\hat{\sigma}^{(n,i)}_{a\rightarrow{\mojet}}}{dx_B\,dQ^2}
=\frac{e^2}{\xi\,x_B^2\,S_H^2}\,\frac{1}{S_n}\,\int 
\left|{\cal M}^{(i)}_{a\rightarrow n}\right|^2\,\mbox{dPS}^{(n)}\,
\left[\delta^{(3)}(\vec{p}_{\mjet}-\hat{p})\,2\,E_{\mjet}\right]\,
\left[\frac{d^{3}p_{\mjet}}{2\,E_{\mjet}}\right]\,,
\end{equation}
where $\hat{p}$ defines the jet momentum in terms of the momenta of the
$n$ partons, $\xi$ is the momentum fraction of the parent hadron carried
by parton $a$, $S_H=(P+l)^2$ is the squared energy of the collision in the 
lepton-proton center of mass, and $x_B$ the usual Bjorken variable. 
The index $i=M,L$ stands for metric and longitudinal contributions 
respectively. $S_n$ is a symmetry factor that account for identical partons 
in the final state.

Since the last two factors between square brackets in equation 
\eqref{eq:totalcs} are Lorentz invariant, 
we can evaluate the latter in the hadronic center of mass while the
former in partonic center of mass. These frames are defined by 
$\vec{P}+\vec{q}=0$ and $\vec{p}_a+\vec{q}=0$, respectively, 
with $P$ and $p_a=\xi\,P$ the momenta of the initial state proton and 
parton respectively. In the hadronic center of mass frame, it is convenient 
to use the transverse energy, $E_\perp$, and pseudo-rapidity $\eta$, while 
in the partonic frame, we use the same variables as for the 
one-particle-inclusive case 
\cite{Daleo:2004pn}:
\begin{align}
y&=-\frac{u}{Q^2+s}&z&=\frac{(Q^2+s)(s+t+u)}
{s\,(Q^2+s+u)}=\frac{s+t+u}{s\,(1-y)}\,,\\
Y&=-\frac{U}{Q^2+S}&Z&=\frac{(Q^2+S)(S+T+U)}
{S\,(Q^2+S+U)}=\frac{S+T+U}{S\,(1-Y)}\,,
\end{align}
with
\begin{align}
s&=(q+p_i)^2 &S&=(q+P)^2\,,\\
t&=-2\,q\cdot p_{\mjet} &T&=-2\,q\cdot p_{\mjet}=t\,,\\
u&=-2\,p_a\cdot p_{\mjet}&U&=-2\,P\cdot p_{\mjet}=u/\xi\,.
\end{align}
Replacing the jet phase space in eq. \eqref{eq:totalcs} we find 
\begin{equation}\label{eq:totalcs2}
\frac{d\hat{\sigma}^{(n)}_{a\rightarrow{\mojet}}}{dx_B\,dQ^2}
=\frac{1}{\xi}\,\int 
\frac{d\,\hat{\sigma}^{(n)}_{a\rightarrow{\mojet}}}{dx_B\,dQ^2\,dy\,dz}\,
\frac{\pi\,E_\perp}{\pi\,\frac{s}{2}\,(1-y)}\,dE_\perp\,d\eta\,,
\end{equation}
where we have introduced 
\begin{equation}\label{eq:pcsfinal}
\frac{d\,\hat{\sigma}^{(n)}_{a\rightarrow{\mojet}}}{dx_B\,dQ^2\,dy\,dz}=
C_n\,\int 
\left|{\cal M}_{a\rightarrow n}\right|^2\,\mbox{dPS}^{(n)}\,
\left[\delta^{(3)}(\vec{p}_{\mjet}-\hat{p})\,2\,E_{\mjet}\right]\,
\left[\pi\,\frac{s}{2}\,(1-y)\right]\,.
\end{equation}
and $C_n=e^2/(x_B^2\,S_H^2\,S_n)$. 
We omit from now on the index $i$, since all the results that follow are 
valid for both metric and longitudinal components. Notice that equation 
\eqref{eq:pcsfinal} is valid in $d=4$ dimensions. 
Convolving the partonic cross section with appropriate parton densities
$f_{a}(\xi)$, the hadronic cross section can be written as
\begin{equation}\label{eq:hadcsfinal}
\frac{1}{2\,E_\perp}\frac{d\sigma_{\mojet}}{dx_B\,dQ^2\,dE_\perp\,d\eta}=
\frac{1}{Q^2+S}\,\frac{1}{1-y}\,\int_{0}^{Z}\frac{dz}{1-z}\,
\frac{f_{a}(\xi)}{\xi}\,
\frac{d\hat{\sigma}^{(n)}_{a\rightarrow\mojet}}{dx_B\,dQ^2\,dy\,dz}\,.
\end{equation}
At variance with the one-particle-inclusive case, the variable
$y$ is completely determined by $E_\perp$ and $\eta$, 
\begin{align}
\label{eq:Y}
y=Y=\frac{E_\perp}{S^{1/2}}\,e^{-\eta}\,,
\end{align}
while $z$ is not fixed.
Equation (\ref{eq:hadcsfinal}) coincides with equation (13) in 
\cite{Daleo:2004pn} when $D_j(\zeta)=\delta(1-\zeta)$.

\subsection{Jet Contributions}

In addition to the jet analog of the one-particle-inclusive
contribution, where the detected final state jet is originated from the 
fragmentation of just one final state parton, while the other partons play 
as spectators and are integrated over, we have to consider additional 
contributions in the full jet cross section. The former will be denoted 
as $\hat{\sigma}_{a\rightarrow i}$, with initial state parton $a$ and 
fragmenting parton $i$. The latter include, on the 
one hand, contributions accounting for the situations in which the jet is 
formed by two partons which are denoted as $\hat{\sigma}_{a\rightarrow ij}$.
On the other hand, we have to subtract configurations in which the cone 
that defines the jet contains two of the final state partons, that
in the one particle inclusive case were nevertheless classified as 
$\hat{\sigma}_{a\rightarrow i}$.  These contributions will be denoted as
 as $\hat{\sigma}_{a\rightarrow i(j)}$.
Discriminated by their partonic content, we have the following 
${\cal{O}}(\alpha_s^2)$ contributions:
\begin{itemize}
\item $q\rightarrow g+g+q$:
\begin{equation}\label{eq:r1}
\begin{split}
\left[d\hat{\sigma}^{(2g)}_{q\rightarrow q}-2\,d\hat{\sigma}_{q\rightarrow q(g)}\right]+
\left[2\,d\hat{\sigma}_{q\rightarrow g}-2\,d\hat{\sigma}_{q\rightarrow g(g)}
-2\,d\hat{\sigma}_{q\rightarrow g(q)}\right]
+2\,d\hat{\sigma}_{q\rightarrow qg}+
d\hat{\sigma}_{q\rightarrow gg}\,,
\end{split}
\end{equation}
\item $q\rightarrow q+q^\prime+\bar{q}^\prime$:
\begin{equation}\label{eq:r2}
\left[d\hat{\sigma}^{(ij)}_{q\rightarrow q}\right]+
\left[d\hat{\sigma}_{q\rightarrow q^\prime}-d\hat{\sigma}_
{q\rightarrow q^\prime(\bar{q}^\prime)}\right]+
\left[d\hat{\sigma}_{q\rightarrow \bar{q}^\prime}-d\hat{\sigma}_
{q\rightarrow \bar{q}^\prime(q^\prime)}\right]+
d\hat{\sigma}_{q\rightarrow q^\prime\bar{q}^\prime}\,,
\end{equation}
\item $q\rightarrow q+q+\bar{q}$:
\begin{equation}\label{eq:r3}
\left[2\,d\hat{\sigma}^{(ii)}_{q\rightarrow q}-2\,d\hat{\sigma}_
{q\rightarrow q(\bar{q})}\right]+
\left[d\hat{\sigma}_{q\rightarrow \bar{q}}-2\,d\hat{\sigma}_
{q\rightarrow \bar{q}(q)}\right]+
2\,d\hat{\sigma}_{q\rightarrow q\bar{q}}\,,
\end{equation}
\item $g\rightarrow g+q+\bar{q}$:
\begin{equation}\label{eq:r4}
\begin{split}
&\left[d\hat{\sigma}_{g\rightarrow g}-d\hat{\sigma}_{g\rightarrow g(q)}-
d\hat{\sigma}_{g\rightarrow g(\bar{q})}\right]+
\left[d\hat{\sigma}_{g\rightarrow q}-d\hat{\sigma}_
{g\rightarrow q(g)}\right]+\left[d\hat{\sigma}_{g\rightarrow \bar{q}}-
d\hat{\sigma}_{g\rightarrow \bar{q}(g)}\right]+d\hat{\sigma}_{g\rightarrow gq}+d\hat{\sigma}_{g\rightarrow g\bar{q}}\,.
\end{split}
\end{equation}
\end{itemize}
where the  $\hat{\sigma}_{a\rightarrow i}$ terms are those already taken 
into account in the one-particle-inclusive cross section, with the 
adequate combinatoric prefactors. 
Notice that cross sections with identical partons in the final state get 
a symmetry factor, so we have to add them to the corresponding matrix 
elements, before factorization. For instance, all the matrix
elements involved in the first and third reaction have to be multiplied by 
$1/2$ due to the presence of two gluons or two identical quarks in the 
final state.

In the small cone approximation it is customary to neglect 
${\cal{O}}(\delta^2)$ contributions in the cross section, 
which means we can approximate the matrix elements associated 
to two partons forming a jet with the corresponding collinear limit. 
This simplifies drastically the calculation
because in this limit the ${\cal{O}}(\alpha_s^2)$ $1\longrightarrow 3$ matrix 
elements factorize into ${\cal{O}}(\alpha_s)$  $1\longrightarrow 2$ matrix 
elements times LO splitting functions.
In the list above, we have skipped terms 
that vanish in this approximation. For example,  in the
second reaction $q\rightarrow q+q^\prime+\bar{q}^\prime$, the matrix 
elements vanish in the limit in which quark $q$ is collinear with any
of $q^\prime$ or $\bar{q}^\prime$.

Restricting ourselves to the limit when partons $j$ and $k$ become
collinear, for both $d\hat{\sigma}_{a\rightarrow j(k)}$ 
and $d\hat{\sigma}_{a\rightarrow jk}$, the corresponding matrix 
elements are given by \cite{Catani:1998nv} ($d=4+\epsilon$):
\begin{equation}\label{eq:collim}
\left|{\cal M}_{a\rightarrow ijk}\right|^2\stackrel{j\parallel k}
{\longrightarrow}
\frac{2}{s_{jk}}\,4\,\pi\,\mu^{-\epsilon}\,\alpha_s\,
\left|{\cal M}_{a\rightarrow iJ}\right|^2\,
\hat{P}_{J\rightarrow jk}(\alpha,\epsilon)\,,
\end{equation}
where the momenta of the parent parton, $J$, is defined in terms of those of partons $j$ and $k$:
\begin{align}
p_{j}^{\mu}&=\alpha\,p_J^{\mu}+k_\perp^{\mu}-\frac{k_\perp^2}{\alpha}\,
\frac{n^{\mu}}{2\,p_J\cdot n}\,,&
p_{k}^{\mu}&=(1-\alpha)\,p_J^{\mu}-k_\perp^{\mu}-\frac{k_\perp^2}{1-\alpha}\,
\frac{n^{\mu}}{2\,p_J\cdot n}\,,
\end{align}
with $p_J\cdot k_\perp=n\cdot k_\perp=0$ and $p_J^2=n^2=0$. Consequently
\begin{equation}\label{eq:pJ}
p_J=\frac{(1-\alpha)\,p_k-\alpha\,p_j}{1-2\,\alpha}-k_\perp.
\end{equation}
Notice that due to the singularities (poles in $\epsilon$) arising
from the $s_{jk}$ factor in the denominator when integrating over the
phase space, we need the splitting functions in \eqref{eq:collim} up to order 
$\epsilon$. The poles in $\alpha=1$ in the diagonal kernels ($J=j$) will 
give rise poles in $z=0$.

\subsection{$d\hat{\sigma}_{a\rightarrow j(k)}$ contributions}
As we mentioned, these contributions correspond to configurations where the 
jet 
is formed solely by parton $j$, but parton $k$ lies within the cone. 
They can be computed along the lines of what has been done in 
reference \cite{Jager:2004jh} but now for DIS kinematics. In the partonic center of 
mass frame we have
\begin{align}
s_{jk}&=2\,p_j\cdot p_k=2\,E_j\,E_k\,(1-\cos\theta_{jk})\,,& 
E_j&=\frac{s^{1/2}}{2}\,(1-z\,(1-y))\,,& E_k&
\simeq \frac{s^{1/2}}{2}\,z\,(1-y)\,,
\end{align}
while the collinear limit implies $\alpha=1-z\,(1-y)$.
Then, we can write the matrix element for $d\hat{\sigma}_{a\rightarrow j(k)}$ as
\begin{equation}\label{eq:collim1}
\begin{split}
\left|{\cal M}_{a\rightarrow ijk}\right|^2\stackrel{j\parallel k}
{\longrightarrow}&
\frac{1-z\,(1-y)}{z\,(1-y)}\,\frac{1}{E_j^2\,(1-\cos\theta_{jk})}
\,
\hat{P}_{J\rightarrow jk}(\alpha
,\epsilon) 
\,4\,\pi\,\mu^{-\epsilon}\,\alpha_s\,
\left|{\cal M}_{a\rightarrow iJ}\left(s,y^\prime=\frac{y}{1-z\,(1-y)}\right)
\right|^2\,.
\end{split}
\end{equation}
For the phase space we have,
\begin{equation}
\begin{split}
\mbox{dPS}^{(3)}&=\frac{\pi\,s}{(4\,\pi)^4}\,
\frac{2^{4-d}}{\Gamma^2\left(\frac{d-2}{2}\right)}\,
\left(\frac{s}{4\,\pi}\right)^{d-4}\,\sin^{d-4}\theta_{jk}\,d\cos\theta_{jk}
(1-y)^2\,z\,
\left((1-y)^3\,y\,z^2\,(1-z)\right)^{(d-4)/2}dy\,dz\,,
\end{split}
\end{equation}
and combining with the matrix elements in \eqref{eq:collim1}, we have 
\begin{equation}
\frac{d\hat{\sigma}_{a\rightarrow j(k)}}{dx_B\,dQ^2\,dy\,dz}
=\frac{1}{S_n}\,\left(\frac{\delta}{2}\right)^{\epsilon}\,
{\cal F}_{a\rightarrow iJ}\,\left[(1-y)^{1+\epsilon}\,z^{\epsilon}\,
\hat{P}_{J\rightarrow jk}(\alpha=1-z\,(1-y),\epsilon)\right]\,,
\end{equation}
where
\begin{equation}
\begin{split}
{\cal F}_{a\rightarrow iJ}&=\frac{e^2}{x_B^2\,S_H^2}\,
\frac{\alpha_s}{(4\,\pi)^2}\,\frac{2}{\epsilon}\,
\frac{1}{\Gamma^2\left(1+\frac{\epsilon}{2}\right)}\,
\frac{\left(y\,(1-y)\,(1-z)\right)^{\epsilon/2}}{1-z\,(1-y)}\,
\left(\frac{s}{4\,\pi\,\mu^2}\right)^{\epsilon}
\times\mu^{\epsilon}\,
\left|{\cal M}_{a\rightarrow iJ}
\left(s,y^\prime=\frac{y}{1-z\,(1-y)}\right)\right|^2\,.
\end{split}
\end{equation}
Notice that with this notation, the variables $y$ and $z$ in 
${\cal F}_{a\rightarrow iJ}(s,y,z)$ are defined in terms
of the second parton in the final state $J$. As mentioned above, the poles 
in $\alpha=1$ in the diagonal kernels, appear as poles in $z=0$ 
(notice that $y=1$ is protected by the $1-y$ factor coming
from the phase space). This poles have to be prescribed as usual, leading
to double poles in $\epsilon$, together with the appearance of $\delta(z)$
and `plus' distributions in $z=0$. The double poles  cancel with terms
coming from $d\sigma_{a\rightarrow jk}$, whereas the remaining $1/\epsilon$ poles cancel 
terms generated when the fragmentation function factorization of the
one-particle-inclusive cross section is undone.

\subsection{$d\hat{\sigma}_{a\rightarrow jk}$ contributions}
These terms account for jets formed by two partons, $j$ and $k$. In the 
collinear limit \eqref{eq:collim} we have:
\begin{equation}
E_{\mjet}=E_J\simeq\frac{E_j}{\alpha}\simeq\frac{E_k}{1-\alpha}\,,
\end{equation}
where $\alpha$ is again the momentum fraction in the splitting function.
The phase space in this case is 
\begin{equation}
\begin{split}
\mbox{dPS}^{(3)}
&=\frac{1}{(4\,\pi)^{3}}\,
\frac{1}{\Gamma^2\left(\frac{d-2}{2}\right)}\,
\left(\frac{s}{4\,\pi}\right)^{d-4}\,\delta(z)\,(y\,(1-y))^{(d-4)/2}
\left(\frac{E_{\mjet}}{\sqrt{s}}\right)^{d-4}\,
\frac{E_{\mjet}^3}{E_k}\,\alpha^{d-3}\,d\alpha\,\sin^{d-3}\theta_j\,
d\theta_j\,dy\,dz\,,
\end{split}
\end{equation}
where we have chosen the frame in such a way that the jet momentum is oriented 
in the $z$ axis. This result exhibits some differences with the previous case. 
Due to the presence of the $\delta(z)$ function, the 
$d\hat{\sigma}_{a\rightarrow jk}$ 
pieces only contribute at $z=0$. On the other hand, now we have to perform 
explicitly the integration over the argument of the splitting functions,
$\alpha$. In addition, we see that the only dependence on the angle in
the matrix element comes, again, form the $s_{jk}$ denominator. However,
now we are integrating over $\theta_j$ and not over $\theta_{jk}$. As in
\cite{Jager:2004jh}, we find
\begin{align}
\theta_j&\simeq (1-\alpha)\,\theta_{jk}\,,&\theta_k&=\alpha\,\theta_{jk}\,.
\end{align}
and the integration limit over $\theta_j$ is given by $\delta$ if
$E_j<E_k$ and by $(1-\alpha)/\alpha\,\delta$ if $E_j>E_k$. Performing the 
angular integrals in the partonic center of mass frame,
\begin{equation}
\frac{d\hat{\sigma}_{a\rightarrow jk}}{dx_B\,dQ^2\,dy\,dz}
=\frac{1}{S_n}\,\left(\frac{\delta}{2}\right)^{\epsilon}\,
{\cal F}_{a\rightarrow iJ}\,
\delta(z)\,\int_{0}^{1}d\alpha\,
\left[\Theta\left(\frac{1}{2}-\alpha\right)\,\alpha^{\epsilon}
+\Theta\left(\alpha-\frac{1}{2}\right)\,(1-\alpha)^{\epsilon}\right]
\hat{P}_{J\rightarrow jk}(\alpha,\epsilon)
\,,
\end{equation}

\subsection{Cancellation of singularities.}
The last step consists in reverting the factorization of final state 
collinear singularities already done in the finite one-particle-inclusive 
cross section. The simplest way to implement this step is simply to add to the
already finite one-particle-inclusive cross section the terms 
we customary factorize into fragmentation functions, with the 
opposite sign. This provides a check of the whole calculation, 
as the simple poles still appearing in the correction terms have to cancel
when we add these contributions.

The terms that have to be added can be read from the following factorization 
prescription formula for the partonic cross sections at first order, 
when renormalizing fragmentation functions:
\begin{equation}\label{eq:facto}
\begin{split}
\frac{d\hat{\sigma}_{a\rightarrow i}}{dx_B\,dQ^2\,dy}
\longrightarrow &
\,\frac{\alpha_s}{2\,\pi}\left(-\frac{2}{\epsilon}
\right)\,\frac{1-y}{1-z\,(1-y)}\,
\frac{\Gamma(1+\epsilon/2)}{\Gamma(1+\epsilon)}
\left(\frac{M_D^2}{4\,\pi\mu^2}\right)^{\epsilon/2}
\frac{d\hat{\sigma}_{a\rightarrow i}}
{dx_B\,dQ^2\,dy}
\sum_{k}\,P_{i\rightarrow jk}(1-z\,(1-y))\,.
\end{split}
\end{equation}
The splitting kernels in \eqref{eq:facto} are the regularized ones in 4 
dimensions.  The Born level cross sections are, in turn, given by:
\begin{equation}\label{eq:ct1}
\begin{split}
\frac{d\hat{\sigma}_{a\rightarrow i}}{dx_B\,dQ^2\,dy}&=
\frac{e^2}{x_B^2\,S_H^2}\,\frac{1}{8\,\pi}\,
\frac{1}{\Gamma(1+\epsilon/2)}\,\left(\frac{s}{4\,\pi\mu^2}\right)^{\epsilon/2}
\,y^{\epsilon/2}\,(1-y)^{\epsilon/2}\,\mu^\epsilon\,\sum_{j}\left|
{\cal M}_{a\rightarrow ij}\right|^2\,.
\end{split}
\end{equation}
The additional factor of $1/\xi$ in the hadronic cross section is simply
re-expressed in terms of the new variables and factors out everywhere. 

Taking into account all the first order cross sections, we have the 
following contributions coming from reverting final state factorization
\begin{equation}\label{eq:ctsum}
\begin{split}
d\hat{\sigma}_{ct}=&d\hat{\sigma}_{q\rightarrow(qg)}+
d\hat{\sigma}_{q\rightarrow(gq)}+d\hat{\sigma}_{q\rightarrow(gg)}+\sum_{q^\prime}\left(
d\hat{\sigma}_{q\rightarrow(q^\prime\bar{q^\prime})}+
d\hat{\sigma}_{q\rightarrow(\bar{q^\prime}q^\prime)}\right) 
\\&
+d\hat{\sigma}_{g\rightarrow(qg)}+d\hat{\sigma}_{g\rightarrow(\bar{q}g)}+
d\hat{\sigma}_{g\rightarrow(gq)}+d\hat{\sigma}_{g\rightarrow(g\bar{q})}\,,
\end{split}
\end{equation}
where we have introduced the notation $d\sigma_{a\rightarrow (jk)}$:
\begin{equation}
\begin{split}
\frac{d\hat{\sigma}_{a\rightarrow (jk)}}{dx_B\,dQ^2\,dy\,dz}=
=\frac{\Gamma^2(1+\epsilon/2)}{\Gamma(1+\epsilon)}\,
\left(\frac{M_D^2}{s}\right)^{\epsilon/2}\,
{\cal F}_{a\rightarrow iJ}
\times(1-y)\,(1-z\,(1-y))^{-\epsilon}\,P_{J\rightarrow jk}(1-z\,(1-y))\,.
\end{split}
\end{equation}
Notice that we have already added a minus sign to these contributions, as
they must be subtracted to recover the original, un-factorized, 
one-particle-inclusive cross section. We also took into account the fact
that at first order, the sum in eq. \eqref{eq:ct1} contains only one term.

As mentioned, doubles poles must cancel explicitly between the 
$\sigma_{a\rightarrow j(k)}$ and $\sigma_{a\rightarrow jk}$ contributions. 
The remaining single poles must cancel with $d\sigma_{a\rightarrow (jk)}$.
To make this cancellations more transparent, it is 
convenient to group the different contributions in eqs. 
\eqref{eq:r1}-\eqref{eq:r4} and the ones in \eqref{eq:ctsum}. 
Starting with the quark initiated reactions and omitting 
$d\sigma_{a\rightarrow j}$ contributions (which are already finite) we have
\begin{equation}
\begin{split}
f_{q}\rightarrow &-\Big[2\,d\hat{\sigma}_{q\rightarrow g(q)}
-d\hat{\sigma}_{q\rightarrow (gq)}\Big]
-n_f\,\Big[
d\hat{\sigma}_{q\rightarrow q^\prime(\bar{q}^\prime)}+
d\hat{\sigma}_{q\rightarrow \bar{q}^\prime(q^\prime)}-
d\hat{\sigma}_{q\rightarrow (q^\prime\bar{q}^\prime)}-
d\hat{\sigma}_{q\rightarrow (\bar{q}^\prime q^\prime)}\Big]\\
&-\Big[2\,d\hat{\sigma}_{q\rightarrow q(g)}-
2\,d\hat{\sigma}_{q\rightarrow qg}
-d\hat{\sigma}_{q\rightarrow (qg)}\Big]
-\Big[2\,d\hat{\sigma}_{q\rightarrow g(g)}
-d\hat{\sigma}_{q\rightarrow gg}
-d\hat{\sigma}_{q\rightarrow (gg)}-\,n_f\,
d\hat{\sigma}_{q\rightarrow q^\prime\bar{q}^\prime}
\Big]\,.
\end{split}
\end{equation}
For the gluon initiated reactions, 
we have:
\begin{equation}
\begin{split}
f_{g}\rightarrow &-\Big[d\hat{\sigma}_{g\rightarrow g(q)}
-d\hat{\sigma}_{g\rightarrow (gq)}\Big]
-\Big[d\hat{\sigma}_{g\rightarrow g(\bar{q})}
-d\hat{\sigma}_{g\rightarrow (g\bar{q})}\Big]
-\Big[{d\hat{\sigma}_{g\rightarrow q(g)}}-
{d\hat{\sigma}_{g\rightarrow qg}}
-d\hat{\sigma}_{g\rightarrow (qg)}\Big]\\
&-\Big[{d\hat{\sigma}_{g\rightarrow \bar{q}(g)}}-
{d\hat{\sigma}_{g\rightarrow \bar{q}g}}
-d\hat{\sigma}_{g\rightarrow (\bar{q}g)}\Big]\,.
\end{split}
\end{equation}
In Appendix I, we list the results corresponding to the cancellations 
for each of the terms in square brackets, which completes the calculation.

\section{Comparison with Monte Carlo results.}

\begin{figure}[hbt]
\includegraphics[width=14cm]{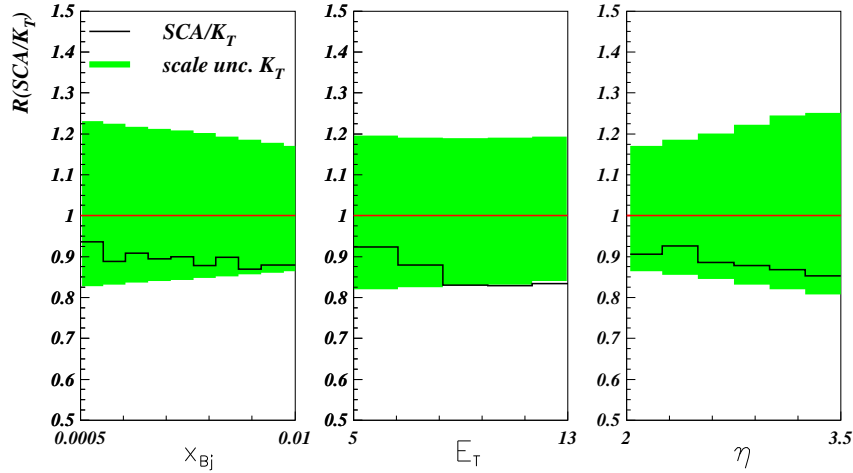}
\caption{Ratio of NLO SCA estimate and the full Monte Carlo prediction with 
$k_T$ jet reconstruction.}\label{fig:CODIS}
\end{figure}

Having obtained the finite expressions for the NLO corrections in the SCA, in
this section we investigate the 
accuracy of the approximation. 
In Figure \ref{fig:CODIS}  we compare the outcome of the SCA for the single 
jet inclusive DIS cross section with a full Monte Carlo NLO calculation of 
\cite{disent}. The jets are reconstructed in the Breit frame and the 
rates between both results are computed in the typical  kinematic range of 
forward jet DIS 
experiments at HERA. Jets are defined using the 
inclusive $k_T$ cluster algorithm \cite{ES} in the Monte Carlo, and 
a cone radius of $R=0.7$ for the SCA value for which the agreement between 
both jet definitions is maximized.
 
 The rates are presented as a function of the 
transverse momentum $E_T$ and rapidity  $\eta$ of the jet, both measured in 
the laboratory frame, and Bjorken momentum fraction $x_{B}$  
\setlength{\unitlength}{1.mm}
In both cases we use the MRST02 NLO set of parton densities \cite{MRST02} and 
we compute $\alpha_s$ at NLO fixing $\Lambda_{QCD}$ as in the MRST analysis
so $\alpha_s(M_Z)=0.1197$.
The rapidity variable varies between $2.0$ and $3.5$, the Bjorken 
variable spans the interval between $0.0005$ and $0.01$, the
transverse momentum of the jet starts at $5\,\mbox{GeV}$, and $Q^2$ ranges from 
$20$ to $100\,\mbox{GeV}^2$. 

As it has been observed in references \cite{SCA2} and \cite{Jager:2004jh},
for hadronic collisions, even for rather large cone radius the SCA 
gives acceptable approximations within less than a ten percent of the full 
Monte Carlo result. This is also the case for DIS and the rates show a very 
mild dependence in the kinematical variables.


Certainly, the accuracy of SCA is the better for smaller cone radius, however 
the error introduced by the approximation with $R=0.7$, which is of the 
order of a 10 \%, always underestimating the cross section and with a very 
mild dependence on the relevant variables,  is comparable or smaller than the 
theoretical uncertainty coming from the particular choice of the factorization 
and renormalization scales, characteristic of the NLO corrections in this 
kinematical region. As we show in the following section,  
this means that we can safely use the SCA results as an estimate of the size
and behavior of the NLO corrections.
For comparison, in the Figure \ref{fig:CODIS}, we also plot as a band the 
uncertainty resulting from varying the factorization and renormalization 
scale by a factor of two in the full Monte Carlo result.

\section{Phenomenological consequences}

Having established our level of confidence in the SCA results, we proceed 
analyzing the distinctive features of the NLO corrections in the forward 
region. 
We do the analysis in the typical kinematic region tested by DESY experiments,
where large higher order effects have been observed.
\setlength{\unitlength}{1.mm}
\begin{figure}[hbt]
\includegraphics[width=14.5cm]{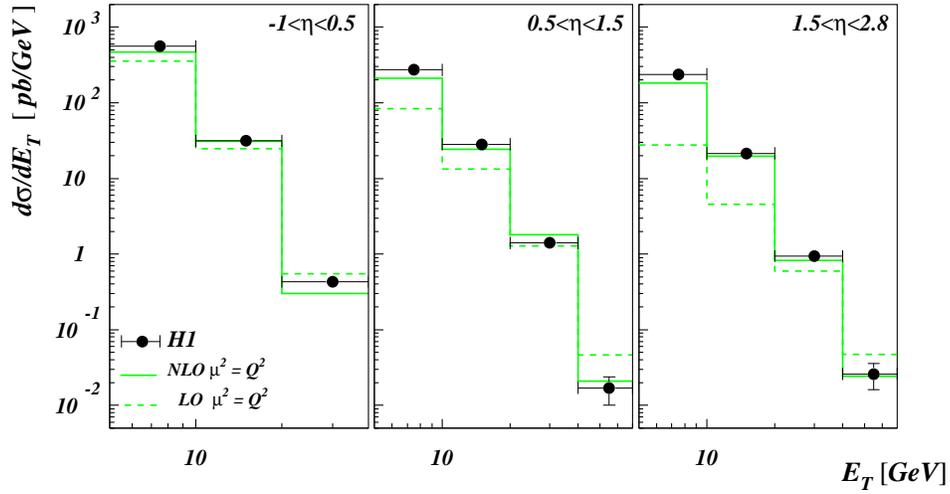}
\caption{Size of NLO corrections as a function of $E_T$ for different rapidity
regions.}\label{fig:H1K}
\end{figure} 

The most striking of these features is the size of the NLO corrections
as the rapidity of the jets increases. In Figure \ref{fig:H1K} 
we show both LO an NLO partonic level expectations coming from the SCA 
approach in three different regions of rapidity for $5<Q^2<100 \mbox{GeV}^2$ and 
$0.2<y_{el}<0.6$ as a function of transverse jet momentum. 
Clearly, NLO corrections, which are moderate for central rapidities, 
become significantly large in the forward region. This feature is due to a 
suppression of the LO contributions rather than to an increase in rapidity 
of the cross sections. K-factors can exceed an
order of magnitude there, invalidating the lowest order approximation. 
We have included for reference the data obtained by H1 in that 
kinematical range \cite{Adloff:2002ew}, although one should keep in mind 
that for a precise comparison hadronization effects should also be taken 
into account.
\setlength{\unitlength}{1.mm}
\begin{figure}[hbt]
\includegraphics[width=14.5cm]{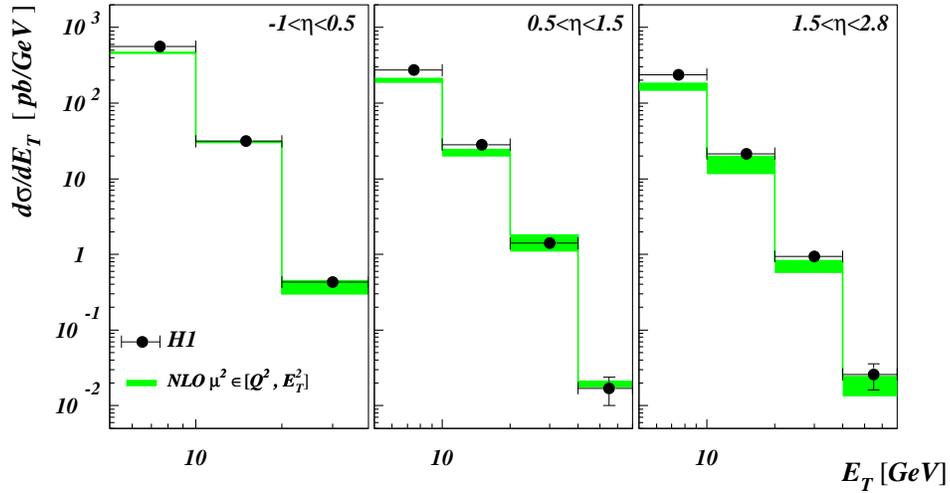}
\caption{Scale uncertainty in NLO corrections as a function of $E_T$ 
for different rapidity
regions.}\label{fig:H1B}
\end{figure} 

Another interesting feature of NLO corrections to be taken into account is 
the rather large uncertainty these corrections show associated with
the choice for the factorization and renormalization scales. In Figure 
 \ref{fig:H1K}  we adopted 
$\mu^2=Q^2$ for the factorization and renormalization scale. The choice
for these scales is in principle arbitrary; the differences found in any 
perturbative estimate coming from some particular choice for the scale or 
other, become smaller as more terms in the perturbation series are included. 
In inclusive DIS $Q^2$ is the typical choice, while $E_T^2$ is the one 
favored in jet physics. In Figure \ref{fig:H1B}  we plot the uncertainty 
bands corresponding to vary the scale from $Q^2$ to $E_T^2$, the two main 
scales of the process under consideration.

With such a large scale uncertainty, any particular choice will probably lead
to miss the data at some point. One possible choice in these cases is to take
the average between them, which leads to an intermediate estimate within the 
band. Again we can see that in the most forward bin, the
uncertainty associated to the choice of the scale becomes more prominent and 
may be as large as a factor of two.
\setlength{\unitlength}{1.mm}
\begin{figure}[hbt]
\includegraphics[width=14cm]{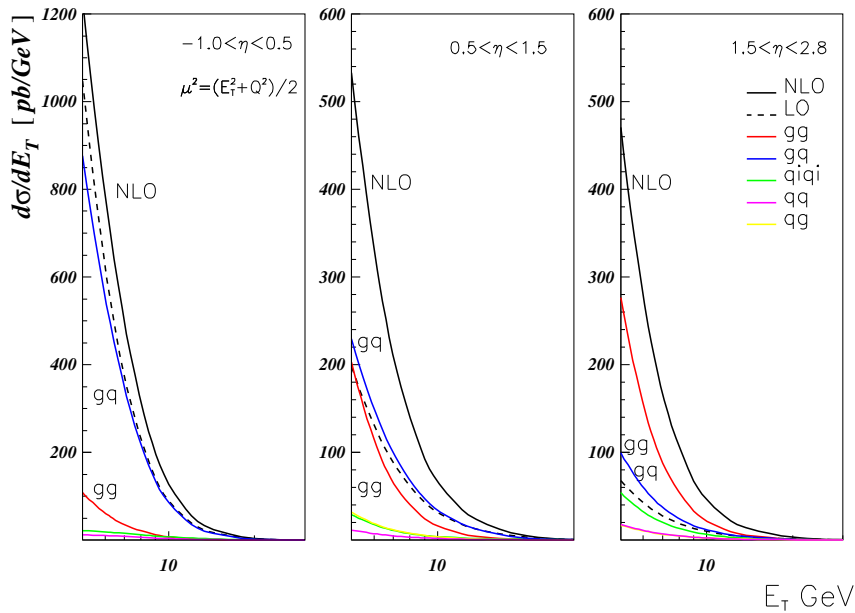}
\caption{Partonic contributions as a function of $E_T$ 
for different rapidity
regions.}\label{fig:H1P}
\end{figure} 

In order to understand the correlation between higher order corrections 
and rapidity, it is useful to discriminate the different 
partonic contributions, classifying them as in the one particle inclusive 
case in terms of the initial state parton, $i$, and the one taken as the seed 
for the jet, $j$, as in $\sigma_{i\rightarrow j}$. The additional 
contributions 
coming from the first two terms in Eq.(32) were added to the 
$\sigma_{g\rightarrow g}$ contribution, the remaining 
terms in that equation were associated to $\sigma_{g\rightarrow q}$. 
Contributions in Eq. (31) were taken together with $\sigma_{q\rightarrow g}$.
     
\begin{figure}[hbt]
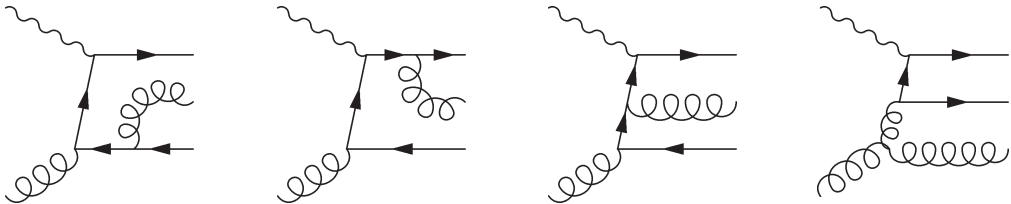

\begin{center}
\begin{minipage}{35mm}
\begin{fmfchar*}(25,25)
  \fmfstraight  \fmfset{arrow_len}{3mm}
  \fmfleft{gi,i1,i2,i3,gamma} 
  \fmfright{o1,aqf,gf,qf,o2} 
  \fmf{photon,width=5}{gamma,vq}
  \fmf{gluon,width=5}{gi,vg1}
  \fmf{fermion,width=5}{aqf,vg2,vg1}
   \fmf{plain,width=5}{vg1,vaux,vq}
  \fmf{phantom_arrow,tension=0}{vg1,vq}
  \fmf{fermion,width=5}{vq,qf}
  \fmffreeze
  \fmf{gluon,left,width=5}{vg2,gf}
\end{fmfchar*}
\end{minipage}
\begin{minipage}{35mm}
\begin{fmfchar*}(25,25)
  \fmfstraight  \fmfset{arrow_len}{3mm}
  \fmfleft{gi,i1,i2,i3,gamma} 
  \fmfright{o1,aqf,gf,qf,o2} 
  \fmf{photon,width=5}{gamma,vq}
  \fmf{gluon,width=5}{gi,vg1}
  \fmf{fermion,tension=0.5,width=5}{aqf,vg1}
   \fmf{plain,width=5}{vg1,vaux,vq}
  \fmf{phantom_arrow,tension=0}{vg1,vq}
  \fmf{fermion,tension=2,width=5}{vq,vg2,qf}
  \fmffreeze
  \fmf{gluon,left,width=5}{gf,vg2}
\end{fmfchar*}
\end{minipage}
\begin{minipage}{35mm}
\begin{fmfchar*}(25,25)
  \fmfstraight  \fmfset{arrow_len}{3mm}
  \fmfleft{gi,i1,i2,i3,gamma} 
  \fmfright{o1,aqf,gf,qf,o2} 
  \fmf{photon,width=5}{gamma,vq}
  \fmf{gluon,width=5}{gi,vg1}
  \fmf{fermion,tension=0.5,width=5}{aqf,vg1}
   \fmf{fermion,width=5}{vg1,vg2,vq}
  \fmf{fermion,width=5}{vq,qf}
  \fmffreeze
  \fmf{gluon,width=5}{vg2,gf}
\end{fmfchar*}
\end{minipage}
\begin{minipage}{35mm}
\begin{fmfchar*}(25,25)
  \fmfstraight  \fmfset{arrow_len}{3mm}
  \fmfleft{gi,i1,i2,i3,gamma} 
  \fmfright{o1,aqf,gf,qf,o2} 
  \fmf{photon,width=5}{gamma,vq}
  \fmf{gluon,width=5}{vg2,vg1,gi}
  \fmf{gluon,tension=0.5,width=5}{vg1,aqf}
   \fmf{fermion,width=5}{vg2,vq,qf}
  \fmffreeze
  \fmf{fermion,width=5}{vg2,gf}
\end{fmfchar*}
\end{minipage}
\end{center}
\caption{Gluon initiated contributions at  ${\cal{O}}(\alpha_s^2)$}
\label{fig:diag}
\end{figure}

In Figure \ref{fig:H1P} we show the different partonic contributions
in the three rapidity regions. While in the central region the LO 
contribution is very close to the full NLO estimate, in the forward region
it is significantly smaller. In the former region, the cross section
is dominated by the $\sigma_{g\rightarrow q}$ contributions 
(an initial state gluon with a quark originating the jet), which are already 
present at LO, while in the latter the dominants are 
$\sigma_{g\rightarrow g}$, which are pure NLO, shown in 
Figure  \ref{fig:diag}. These contributions start at 
order $\alpha_s^2$ so the NLO result is its lowest order estimate. The reason
for their dominance over the LO is just the kinematical region chosen which 
suppress LO configurations. The dominance of the $\sigma_{g\rightarrow g}$ 
over $\sigma_{g\rightarrow q}$  NLO contributions can be traced back to the 
negative 'plus' contributions which are proportional $C_F^2$ in the case of 
the former and to $C_F C_A$ for the latter. 

Since the dominant partonic process in the forward region is accounted, at 
order $\alpha_s^2$, only by only its lowest order contribution, it 
is effectively a LO estimate and most probably receives significant 
higher order corrections.
The first order corrections for other partonic processes in this kinematic 
region rise typically to 50\% effects, so it would not be surprising that
the NLO estimate falls short of the data, specially if a more stringent
kinematic range is explored.
\setlength{\unitlength}{1.mm}
\begin{figure}[hbt]
\includegraphics[width=13cm]{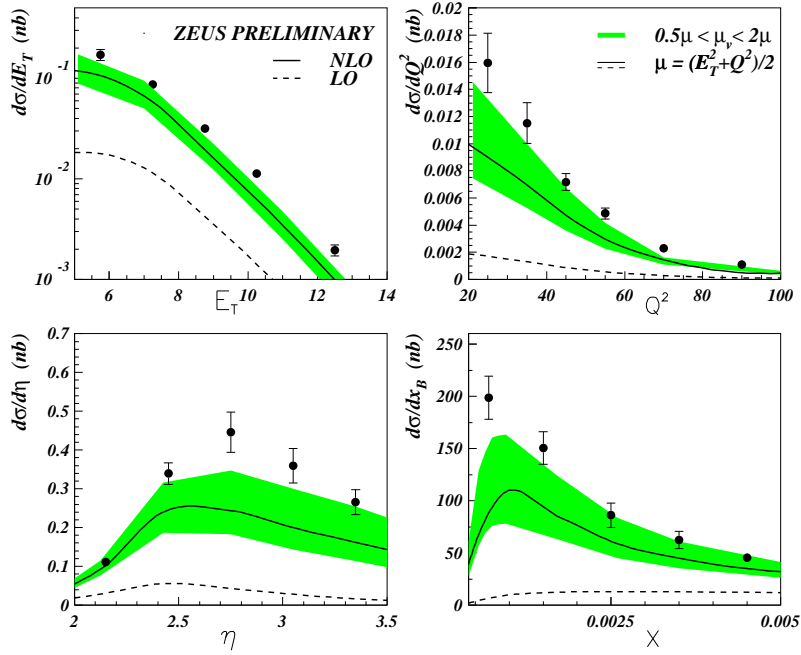}
\caption{
NLO estimates against ZEUS very forward preliminary data \cite{zeusdis}}
\label{fig:zeus}
\end{figure} 
\setlength{\unitlength}{1.mm}
\begin{figure}[hbt]
\includegraphics[width=12.cm]{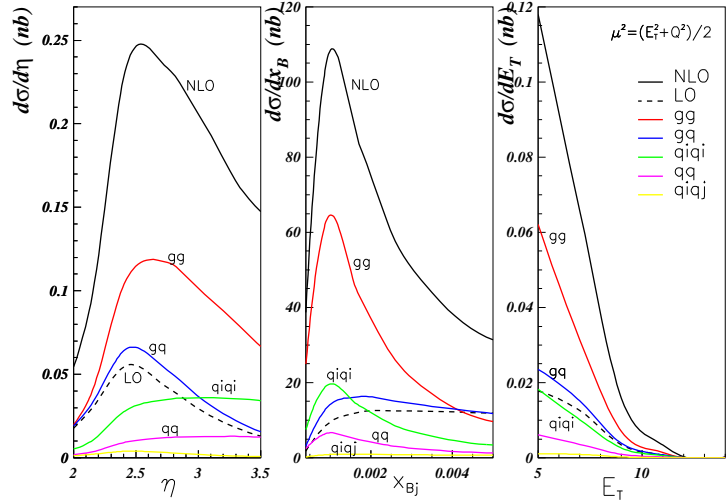}
\caption{Partonic contributions in Zeus very forward measurement
\cite{zeusdis}}
\label{fig:proczeus}
\end{figure} 
 This is precisely what ZEUS has reported in preliminary analyses of 
measurements in the very forward region \cite{zeusdis}.  

In Figure \ref{fig:zeus} we
plot the NLO estimates for the cross section as distributions in different 
variables together with ZEUS preliminary data \cite{zeusdis}  
. The estimate correspond to rapidities between 
$2.0$ and $3.5$, the Bjorken variable in the interval $0.0004$ and $0.005$, 
the transverse momentum of the jet starting at $5\,\mbox{GeV}$, and the 
virtuality of the photon $Q^2$ range from $20$ to $100\,\mbox{GeV}^2$.  
The NLO estimate falls short of the preliminary data, and only allowing a 
rather large scale uncertainty it may be considered consistent with the 
measurements, specially at small $x_B$.

\begin{figure}[hbt]
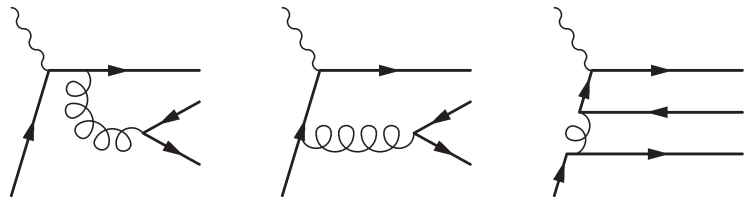

\begin{center}
\begin{minipage}{35mm}
\begin{fmfchar*}(25,25)
  \fmfstraight  \fmfset{arrow_len}{3mm}
  \fmfleft{qi,i1,i2,i3,gamma} 
  \fmfright{g0,g1,g2,g3,qf,o1,o2} 
  \fmf{photon,tension=2,width=5}{gamma,vq}
  \fmf{plain,tension=4}{qi,b1,b2,b3,vq}
  \fmf{fermion,tension=0}{qi,vq}
  \fmf{plain,tension=3}{vq,a1}
  \fmf{fermion,tension=2}{a1,a2}
  \fmf{plain,tension=2}{a2,qf}
  \fmffreeze
  \fmf{phantom}{b2,gi}
  \fmf{gluon,left,width=5,tension=0}{gi,a1}
  \fmf{quark}{g3,gi}
  \fmf{quark}{gi,g1}
\end{fmfchar*}
\end{minipage}
\begin{minipage}{35mm}
\begin{fmfchar*}(25,25)
  \fmfstraight  \fmfset{arrow_len}{3mm}
  \fmfleft{qi,i1,i2,i3,gamma} 
  \fmfright{g0,g1,g2,g3,qf,o1,o2} 
  \fmf{photon,tension=2,width=5}{gamma,vq}
  \fmf{plain,tension=4}{qi,b1,b2,b3,vq}
  \fmf{fermion,tension=0}{qi,vq}
  \fmf{plain,tension=3}{vq,a1}
  \fmf{fermion,tension=2}{a1,a2}
  \fmf{plain,tension=2}{a2,qf}
  \fmffreeze
  \fmf{gluon,width=5}{b2,gi}
  \fmf{quark}{g3,gi}
  \fmf{quark}{gi,g1}
\end{fmfchar*}
\end{minipage}
\begin{minipage}{35mm}
\begin{fmfchar*}(25,25)
  \fmfstraight  \fmfset{arrow_len}{3mm}
  \fmfleft{qi,i1,i2,i3,gamma} 
  \fmfright{g0,g1,g2,g3,g4,g5,qf,o1,o2,o3} 
  \fmf{photon,tension=2,width=5}{gamma,vq}
  \fmf{phantom,tension=6}{qi,b1,b2,b3,b4,b5,vq}
  \fmf{fermion,tension=0}{qi,b2}
  \fmf{fermion,tension=0}{b4,vq}
  \fmf{gluon,tension=0,width=5}{b2,b4}
  \fmf{plain,tension=3}{vq,a1}
  \fmf{fermion,tension=2}{a1,a2}
  \fmf{plain,tension=2}{a2,qf}
  \fmffreeze
  \fmf{quark}{b2,g2}
  \fmf{quark}{g4,b4}
\end{fmfchar*}
\end{minipage}
\end{center}
\caption{Typical quark initiated contributions at ${\cal{O}}(\alpha_s^2)$}
\label{fig:diagq}
\end{figure}

Further insight is obtained analyzing the different partonic contributions
as a function of $\eta$ and $x_{B}$.
In Figure \ref{fig:proczeus} it can be noticed that the 
$\sigma_{g\rightarrow g}$ contributions dominate the cross section, specially
at low $x_{B}$ where the gluon parton density grows dramatically and in the 
middle of the rapidity range. In these two regions one can expect the first
order corrections to these processes, starting at NNLO, to be  
significant. In fact, it is there where the NLO estimate can be more distant 
to the data with a particular choice for the scale, as can be seen when 
comparing with the preliminary data in Ref. \cite{zeusdis}. 
At larger $x_{B}$ and $\eta$, $\sigma_{g\rightarrow g}$ 
contributions decrease, even below the LO contribution, but the other NLO 
contributions keep the total NLO effect very large. Of these contributions, 
the most prominent is $\sigma^{(ij)}_{q\rightarrow q}$, that corresponds to 
diagrams, like the ones in Figure \ref{fig:diagq}, where the two quarks 
lines have different flavours, but where the quark that initiates the jet 
has the same flavour as the one in the initial state.

\begin{figure}[hbt]
\includegraphics[width=17cm]{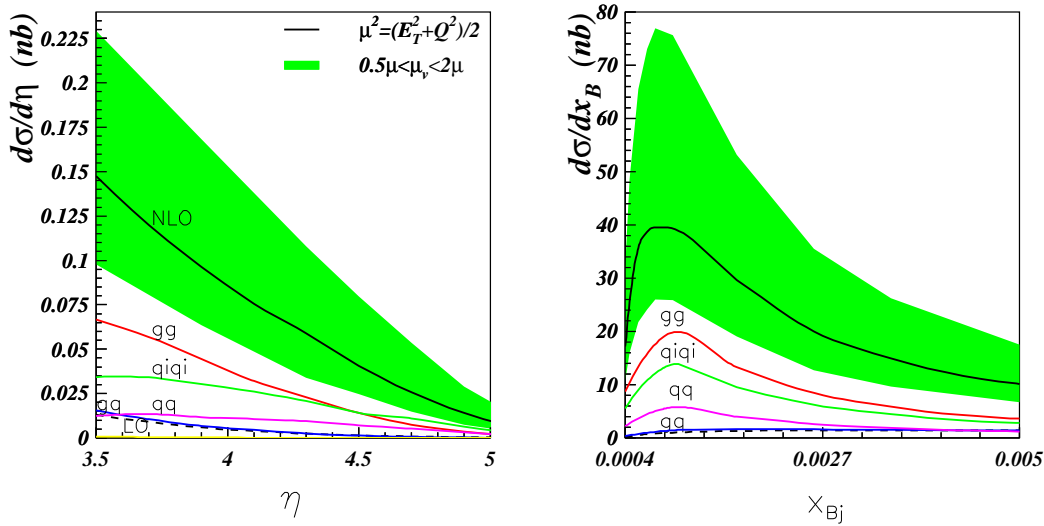}
\caption{
Partonic contributions in an extremely forward region
}\label{fig:etaextremo}
\end{figure}

Going to even higher rapidities, these last contributions eventually dominate 
the 
cross section, with the LO estimate being completely suppressed as shown
in Figure \ref{fig:etaextremo}. In this region, with the two dominant 
contributions being computed at the lowest order, the scale uncertainty of
the NLO estimate is twice as large as that found in the rapidity region of 
Figure \ref{fig:proczeus}.
\setlength{\unitlength}{1.mm}
   
\section{Conclusions}

We have computed the single jet inclusive deep inelastic scattering 
cross section at ${\cal{O}}(\alpha_s^2)$ in the small cone 
approximation.
We found that this approach approximates the full NLO Monte Carlo results 
within a 10\% accuracy, error which is fairly moderate compared to 
the main source of theoretical uncertainty, the scale dependence.

As in the case of hadroproduction in deep inelastic scattering, 
we have found that the dominant partonic processes in very forward jet 
production start at order $\alpha_s^2$, being effectively a lowest order 
estimate. As in any lowest order calculation, there is a large factorization 
scale uncertainty which can not be neglected, and it is likely that there will
be large 
corrections at the subsequent order in perturbation. 
Although taking into account this large dependence on the choice for the 
scale, one can bring agreement between data and NLO estimates, the
difference between them for a particular choice is maximal precisely 
where the partonic contributions computed for the first time are dominant.
This feature is expected to be  even more apparent at higher rapidities, 
and the corresponding measurements will constitute an obligatory benchmark
for the study of QCD at NNLO.

\section{Acknowledgments.}

We warmly acknowledge D. de Florian, C. A. Garc\'{\i}a Canal, C. Glasman and 
Juan Terr\'on for comments and suggestions. We also thank C.Glasman for 
providing us ZEUS data. 
The work of A.D. was supported in part by the 
Swiss National Science Foundation (SNF) through grant No. 200020-109162 and
by the Forschungskredit der Universit\"at Z\"urich.

\section{Appendix}

Here we list the finite results obtained for each of the terms in square
brackets in eqs. (31)  and (32).

\begin{equation}
\begin{split}
T_{q,1}&=-{\cal F}_{q\rightarrow gq}\,\Bigg\{
\left(\frac{\delta}{2}\right)^{\epsilon}\,
\left[(1-y)^{1+\epsilon}\,z^{\epsilon}\,
\hat{P}_{q\rightarrow gq}(1-z\,(1-y))\right]\\
&-\frac{\Gamma(1+\epsilon/2)}{\Gamma(1+\epsilon)}
\left(\frac{M_D^2}{s}\right)^{\epsilon/2}\,(1-y)\,(1-z\,(1-y))^
{-\epsilon}\,P_{q\rightarrow gq}(1-z\,(1-y))
\Bigg\}\,,
\end{split}
\end{equation}
\begin{equation}
\begin{split}
T_{q,2}&=-2\,n_f\,{\cal F}_{q\rightarrow qg}\,\Bigg\{
\left(\frac{\delta}{2}\right)^{\epsilon}\,
\left[(1-y)^{1+\epsilon}\,z^{\epsilon}\,
\hat{P}_{g\rightarrow q\bar{q}}(1-z\,(1-y))\right]\\
&-\frac{\Gamma(1+\epsilon/2)}{\Gamma(1+\epsilon)}
\left(\frac{M_D^2}{s}\right)^{\epsilon/2}\,(1-y)\,(1-z\,(1-y))^
{-\epsilon}\,P_{g\rightarrow q\bar{q}}(1-z\,(1-y))
\Bigg\}\,,
\end{split}
\end{equation}
\begin{equation}
\begin{split}
T_{q,3}&=-{\cal F}_{q\rightarrow gq}\,\Bigg\{
\left(\frac{\delta}{2}\right)^{\epsilon}\,
\bigg[(1-y)^{1+\epsilon}\,z^{\epsilon}\,
\hat{P}_{q\rightarrow qg}(1-z\,(1-y))\\&
-\delta(z)\,\int_{0}^{1}d\alpha\,
G(\alpha,\epsilon)\,\hat{P}_{q\rightarrow qg}(\alpha,\epsilon)
\bigg]\\
&-\frac{\Gamma(1+\epsilon/2)}{\Gamma(1+\epsilon)}
\left(\frac{M_D^2}{s}\right)^{\epsilon/2}\,(1-y)\,(1-z\,(1-y))^
{-\epsilon}\,P_{q\rightarrow qg}(1-z\,(1-y))
\Bigg\}\,,
\end{split}
\end{equation}
\begin{equation}
\begin{split}
T_{q,4}&=-{\cal F}_{q\rightarrow qg}\,\Bigg\{
\left(\frac{\delta}{2}\right)^{\epsilon}\,
\bigg[(1-y)^{1+\epsilon}\,z^{\epsilon}\,
\hat{P}_{g\rightarrow gg}(1-z\,(1-y))\\
&-\frac{\delta(z)}{2}\,\int_{0}^{1}d\alpha\,
G(\alpha,\epsilon)\,\left(\hat{P}_{g\rightarrow gg}(\alpha,\epsilon)
+2\,n_f\,\hat{P}_{g\rightarrow q\bar{q}}(\alpha,\epsilon)\right)
\bigg]\\
&-\frac{\Gamma(1+\epsilon/2)}{\Gamma(1+\epsilon)}
\left(\frac{M_D^2}{s}\right)^{\epsilon/2}\,(1-y)\,(1-z\,(1-y))^
{-\epsilon}\,P_{g\rightarrow gg}(1-z\,(1-y))
\Bigg\}\,.
\end{split}
\end{equation} 

\begin{equation}
\begin{split}
T_{g,1}&=-{\cal F}_{g\rightarrow \bar{q}q}\,\Bigg\{
\left(\frac{\delta}{2}\right)^{\epsilon}\,
\left[(1-y)^{1+\epsilon}\,z^{\epsilon}\,
\hat{P}_{q\rightarrow gq}(1-z\,(1-y))\right]\\
&-\frac{\Gamma(1+\epsilon/2)}{\Gamma(1+\epsilon)}
\left(\frac{M_D^2}{s}\right)^{\epsilon/2}\,(1-y)\,(1-z\,(1-y))^
{-\epsilon}\,P_{q\rightarrow gq}(1-z\,(1-y))
\Bigg\}\,,
\end{split}
\end{equation}
\begin{equation}
\begin{split}
T_{g,2}&=-{\cal F}_{g\rightarrow q\bar{q}}\,\Bigg\{
\left(\frac{\delta}{2}\right)^{\epsilon}\,
\left[(1-y)^{1+\epsilon}\,z^{\epsilon}\,
\hat{P}_{q\rightarrow gq}(1-z\,(1-y))\right]\\
&-\frac{\Gamma(1+\epsilon/2)}{\Gamma(1+\epsilon)}
\left(\frac{M_D^2}{s}\right)^{\epsilon/2}\,(1-y)\,(1-z\,(1-y))^
{-\epsilon}\,P_{q\rightarrow gq}(1-z\,(1-y))
\Bigg\}\,,
\end{split}
\end{equation}
\begin{equation}
\begin{split}
T_{g,3}&=-{\cal F}_{g\rightarrow \bar{q}q}\,\Bigg\{
\left(\frac{\delta}{2}\right)^{\epsilon}\,
\bigg[(1-y)^{1+\epsilon}\,z^{\epsilon}\,
\hat{P}_{q\rightarrow qg}(1-z\,(1-y))\\&
-\delta(z)\,\int_{0}^{1}d\alpha\,
G(\alpha,\epsilon)\,\hat{P}_{q\rightarrow qg}(\alpha,\epsilon)
\bigg]\\
&-\frac{\Gamma(1+\epsilon/2)}{\Gamma(1+\epsilon)}
\left(\frac{M_D^2}{s}\right)^{\epsilon/2}\,(1-y)\,(1-z\,(1-y))^
{-\epsilon}\,P_{q\rightarrow qg}(1-z\,(1-y))
\Bigg\}\,,
\end{split}
\end{equation}
\begin{equation}
\begin{split}
T_{g,4}&=-{\cal F}_{g\rightarrow q\bar{q}}\,\Bigg\{
\left(\frac{\delta}{2}\right)^{\epsilon}\,
\bigg[(1-y)^{1+\epsilon}\,z^{\epsilon}\,
\hat{P}_{q\rightarrow qg}(1-z\,(1-y))\\
&-\delta(z)\,\int_{0}^{1}d\alpha\,
G(\alpha,\epsilon)\,\hat{P}_{q\rightarrow qg}(\alpha,\epsilon)
\bigg]\\
&-\frac{\Gamma(1+\epsilon/2)}{\Gamma(1+\epsilon)}
\left(\frac{M_D^2}{s}\right)^{\epsilon/2}\,(1-y)\,(1-z\,(1-y))^
{-\epsilon}\,P_{q\rightarrow qg}(1-z\,(1-y))
\Bigg\}\,.
\end{split}
\end{equation}

\end{fmffile}
\end{document}